\documentclass[journal]{IEEEtran}

\usepackage{tabularx} 
\pdfoutput=1

\usepackage{cite}

\usepackage{hyperref}       
\usepackage{url}            
\usepackage{booktabs}       
\usepackage{amsfonts}       
\usepackage{nicefrac}       
\usepackage{microtype}      
\usepackage{adjustbox}
\usepackage{hyperref}

\usepackage{bm}
\usepackage[caption=false,font=normalsize, labelfont=sf, textfont=sf]{subfig}
\usepackage[english]{babel}

\usepackage{multirow, rotating, array}
\usepackage{amsfonts, amssymb, amsmath}
\usepackage[capitalise]{cleveref}

\usepackage{floatrow}
\usepackage{flafter}

\usepackage{color, float}
\usepackage{afterpage}
\usepackage{flafter}

\usepackage{algorithmic}
\usepackage[algoruled]{algorithm2e}

\usepackage{wrapfig}
\usepackage{makeidx}
\usepackage{makecell}

\hypersetup{
	colorlinks   = true, %
	urlcolor     = blue, %
	linkcolor    = red, %
	citecolor    = blue   %
}

\begin{document}



\bstctlcite{IEEEexample:BSTcontrol}

\title {Combined analysis of coronary arteries and the left ventricular myocardium in cardiac CT angiography for detection of patients with functionally significant stenosis}

\author{
	Majd~Zreik, Tim~Leiner, Nadieh~Khalili, Robbert~W.~van~Hamersvelt, \\ Jelmer~M.~Wolterink, Michiel Voskuil, Max~A.~Viergever, Ivana~I\v{s}gum%
	\thanks{M.~Zreik and N.~Khalili are with the Image Sciences Institute, University Medical Center Utrecht, The Netherlands (e-mail: m.zreik@umcutrecht.nl).}%
	\thanks{T.~Leiner is with the Department of Radiology, University Medical Center Utrecht and Utrecht University, The Netherlands.}
	\thanks{R.~W.~van~Hamersvelt is with the Department of Radiology, University Medical Center Utrecht, The Netherlands.}%
	\thanks{J.~M.~Wolterink is with the Image Sciences Institute, University Medical Center Utrecht, The Netherlands and the Department of Biomedical Engineering and Physics, Amsterdam University Medical Center.}
	\thanks{M.~Voskuil is with the Department of Cardiology, University Medical Center Utrecht and Utrecht University, The Netherlands.}
	\thanks{M. A.~Viergever is with the Image Sciences Institute, University Medical Center Utrecht and Utrecht University, The Netherlands.}
	\thanks{I.~I\v{s}gum is with the Image Sciences Institute, University Medical Center Utrecht, The Netherlands, the Department of Biomedical Engineering and Physics, Amsterdam University Medical Center, and the Department of Radiology and Nuclear Medicine, Amsterdam University Medical Center.}
	\thanks{This study was financially supported by the project FSCAD, funded by the Netherlands Organization for Health Research and Development (ZonMw) in the framework of the research programme IMDI (Innovative Medical Devices Initiative); project 104003009.}
	\thanks{This work has been submitted to the IEEE for possible publication. Copyright may be transferred without notice, after which this version may no longer be accessible.}}

\markboth{}{}

\maketitle
	

\begin{abstract}
	
	Treatment of patients with obstructive coronary artery disease is guided by the functional significance of a coronary artery stenosis. Fractional flow reserve (FFR), measured during invasive coronary angiography (ICA), is considered the gold standard to define the functional significance of a coronary stenosis.
	Here, we present a method for non-invasive detection of patients with functionally significant coronary artery stenosis, combining analysis of the coronary artery tree and the left ventricular (LV) myocardium in cardiac CT angiography (CCTA) images. 
	We retrospectively collected CCTA scans of 126 patients who underwent invasive FFR measurements, to determine the functional significance of coronary stenoses. 
	We combine our previous works for the analysis of the complete coronary artery tree and the LV myocardium: Coronary arteries are encoded by two disjoint convolutional autoencoders (CAEs) and the LV myocardium is characterized by a convolutional neural network (CNN) and a CAE. Thereafter, using the extracted encodings of all coronary arteries and the LV myocardium, patients are classified according to the presence of functionally significant stenosis, as defined by the invasively measured FFR.
	To handle the varying number of coronary arteries in a patient, the classification is formulated as a multiple instance learning problem and is performed using an attention-based neural network.
	Cross-validation experiments resulted in an average area under the receiver operating characteristic curve of $0.74 \pm 0.01$, and showed that the proposed combined analysis outperformed the analysis of the coronary arteries or the LV myocardium only.
	The results demonstrate the feasibility of combining the analyses of the complete coronary artery tree and the LV myocardium in CCTA images for the detection of patients with functionally significant stenosis in coronary arteries. This may lead to a reduction in the number of unnecessary ICA procedures in patients with suspected obstructive CAD.

\end{abstract}



\section{Introduction}

Among cardiovascular diseases, obstructive coronary artery disease (CAD) is the most common \cite{benjamin2018heart}. Obstructive CAD is defined as a narrowing of the coronary artery lumen, i.e. coronary stenosis, due to a build up of atherosclerotic plaque in the coronary artery wall \cite{cury2016cad}.
A functionally significant coronary stenosis limits blood supply and therefore leads to ischemia and irreversible damage to the left ventricular (LV) myocardium \cite{Pijl96}. However, not all coronary stenoses are functionally significant \cite{cury2016cad}. To decrease CAD morbidity, only functionally significant stenoses require invasive intervention \cite{Pijl96,Toni09,Pijl10,Nune15}. Therefore, to guide invasive intervention, it is important to define the functional significance of a coronary stenosis.

Most often, coronary artery stenosis is noninvasively and visually detected in patients with suspected CAD using cardiac CT angiography (CCTA) scans \cite{Budo08a}. 
However, CCTA has low specificity in defining the functional significance of a coronary stenosis with intermediate severity \cite{Meij08, Bamb11,Ko12}. Therefore, invasive coronary angiography (ICA) is performed on patients with intermediate severity coronary stenosis to measure the fractional flow reserve (FFR) in the coronary arteries. To day, FFR is the gold standard for the functional significance of a coronary stenosis \cite{Pijl96,Toni09}. However, as patients are refereed to ICA based on CCTA with low specificity, coronary artery stenoses in up to 50\% of these patients are eventually determined as functionally non-significant, i.e. these patients unnecessarily undergo invasive FFR measurement \cite{Ko12}.

To reduce the number of unnecessary ICA procedures and invasive FFR measurements, the analysis of CCTA images for defining the functional significance of a coronary artery stenosis has been an active field of research. 
The most validated methods for determining the functional significance of coronary artery stenosis are those that analyze the blood flow in the coronary arteries \cite{Tayl13, Itu12, Nick15,itu2016machine}.
Taylor et al. \cite{Tayl13,Norg14} were the first to employ computational fluid dynamics, in the form of Navier-Stokes equations, to estimate FFR values along the coronary artery. Itu et al. \cite{Itu12} used a parametric lumped heart model to model blood flow with a patient-specific hemodynamics in both healthy and stenotic coronary arteries to estimate its FFR values.
Nickisch et al. \cite{Nick15} employed an electrical patient-specific parametric lumped model to simulate blood flow and pressure along the coronary artery arteries. Later, Itu et al. \cite{itu2016machine} followed a machine-learning-based approach, trained on a large and diverse number of synthetically generated coronary arteries, to compute FFR along the arteries.
Although these flow-based methods \cite{Tayl13,Itu12,Nick15,itu2016machine} achieved good results, they rely on the highly challenging task of accurate segmentation of the coronary artery lumen. To achieve accurate segmentation, these methods require a substantial manual interaction \cite{tesche2017coronary}, which is a very time consuming task, especially in patients with severe atherosclerotic calcifications, coronary stents, and image artifacts \cite{Kiri13a}.

Recently, methods that only analyze the LV myocardium in CCTA images have emerged. In our previous work, we have presented a deep learning-based analysis of the LV myocardium to detect patients with a functionally significant coronary artery stenosis \cite{zreik2018deep,van2018deep}. The LV myocardium is first segmented using a convolutional neural network (CNN) and then characterized by a convolutional autoencoder (CAE). A support vector machine (SVM) is used with the extracted characteristics to classify patients according to the presence of functionally significant stenosis.
Han et al. \cite{Han17} utilized the method presented by Xiong et al. \cite{Xion15} to detect patients with functionally significant stenosis. First, the LV myocardium is segmented and aligned with the 17-segments perfusion model \cite{Cerq02a}. Then, hand-crafted features of each myocardial segment are extracted and employed with an AdaBoost classifier to identify patients with functionally significant stenosis. 

Here, we present a method to automatically detect patients with functionally significant coronary artery stenosis, based on a combined analysis of the complete coronary artery tree and the LV myocardium. First, exploiting our previous work, each coronary artery is analyzed separately through encoding its multi-planar reformatted (MPR) volume into a fixed number of encodings using two disjoint CAEs \cite{zreik2019deep}. Next, exploiting our previously developed method, the LV myocardium is segmented by a CNN and thereafter characterized by a CAE \cite{zreik2018deep}. Finally, based on the extracted encodings of the arteries in the coronary artery tree and the LV myocardium, patients are classified according to the presence of functionally significant stenosis, as defined by the invasively measured FFR.
To handle the varying number of analyzed coronary arteries, the classification is formulated as a multiple instance learning (MIL) problem \cite{ilse2018attention}. Following this approach, the encoding of each coronary artery in the coronary tree, along with the encoding of the LV myocardium, is defined as a single instance in a bag of instances. Then, all instances are combined into a single representation using a trainable attention-based neural network that calculates a learned weighted average of all instances. Finally, the combined representation is used to classify a patient. The contributions of this paper are as follows. First, to the best of our knowledge, this is the first method to combine information derived from both the coronary artery tree and the LV myocardium to detect functionally significant stenosis. Second, to the best of our knowledge, formulating the analysis of all coronary arteries in the coronary tree as a MIL problem has not been presented before.

The remainder of the manuscript is organized as follows. \cref{data_art_myo} describes the data and reference standard. \cref{method_art_myo} describes the method. \cref{results_art_myo} reports our experimental results, which are then discussed in \cref{discussion_art_myo}.

\section{Data} \label{data_art_myo}

This study includes retrospectively collected CCTA scans and invasively measured FFR values of 126 patients (age: $58.9 \pm 9.0$ years, 97 males) acquired between 2012 and 2016. The Institutional Ethical Review Board waived the need for informed consent. 

\subsection{CCTA imaging}

All CCTA scans were acquired using an ECG-triggered step-and-shoot protocol on a 256-detector row scanner (Philips Brilliance iCT, Philips Medical, Best, The Netherlands). A tube voltage of 120 kVp and tube current between 210 and 300 mAs were used. For patients weighing $\le80$ kg contrast medium was injected using a flow rate of 6 mL/s for a total of 70 mL iopromide (Ultravist 300 mg I/mL, Bayer Healthcare, Berlin, Germany), followed by a 50 mL mixed contrast medium and saline (50:50) flush, and next a 30 mL saline flush. For patients weighing $>80$ kg the flow rate was 6.7 mL/s and the volumes of the boluses were 80, 67 and 40 mL, respectively. Images were reconstructed to an in-plane resolution ranging from 0.38 to 0.56 mm, and 0.9 mm thick slices with 0.45 mm spacing.

In each CCTA scan, all visible coronary arteries were automatically tracked and their centerlines were extracted using the method previously described by Wolterink et al. \cite{wolterink2019coronary}. The method tracks the visible coronary arteries, where the arterial centerlines are extracted between the ostia and the most distal visible locations. Using the extracted centerlines, a 3D straightened MPR volumes with 0.3 $mm^3$ isotropic resolution were reconstructed for all coronary arteries and used for further analysis.  In total, in the CCTAs of 126 patients, 2340 arteries were extracted, with $18.5 \pm 4.3$ (interquartile range: 16.0-22.0) arteries per patient. Because the tracking was done from the ostia to the distal parts of the arteries, proximal parts of different arteries may be overlapping.

\subsection{FFR measurements}

\begin{figure}
	\centering
	\includegraphics[width=1.\linewidth]{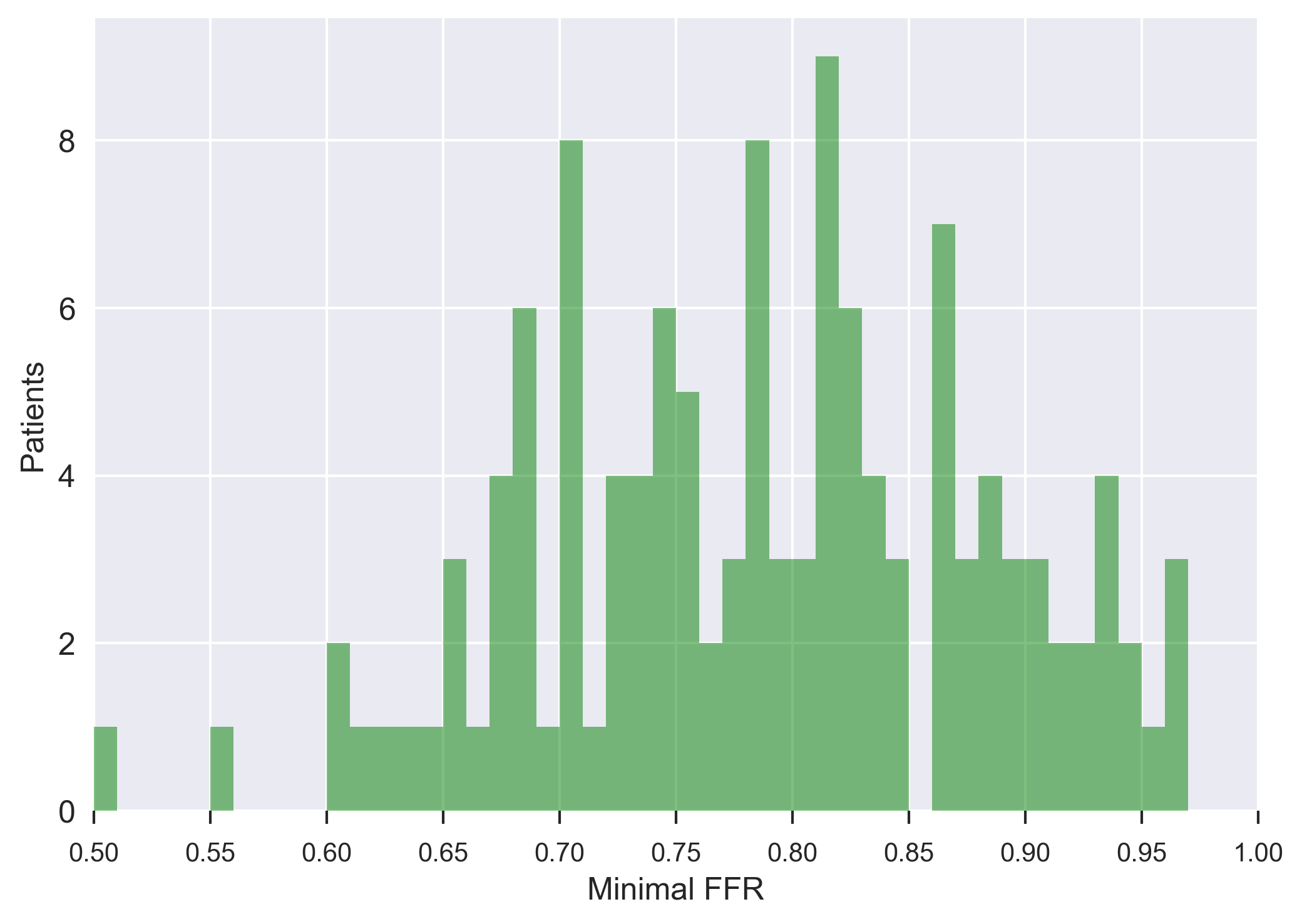}
	\caption{The distribution of the minimal FFR values measured in 126 patients included in this study
	}
	\label{fig:ffr_hist_art_myo}
	
\end{figure}

All patients underwent invasive FFR measurements, up to one year after the acquisition of the CCTA scan. FFR was recorded with a coronary pressure guidewire (Certus Pressure Wire, St. Jude Medical, St. Paul, Minnesota) at maximal hyperemia conditions. Maximal hyperemia was induced by administration of intravenous adenosine (at a rate of 140 $\mu$g/kg per minute) through a central vein. The FFR wire was placed at the most distal part possible in the target artery. Using manual pullback, a single minimal FFR value was assessed and recorded for each artery. When multiple FFR measurements were available, or measurements for multiple stenoses were available, the minimum value was taken as the standard of reference for the patient. The average of the minimal invasively measured FFR values in the 126 patients is $0.79 \pm 0.10$ (interquartile range: 0.72-0.86), where the distribution of minimal FFR values is shown in \cref{fig:ffr_hist_art_myo}.

\section{Methods}\label{method_art_myo}

\begin{figure*}
	\centering
	\includegraphics[width=1\linewidth]{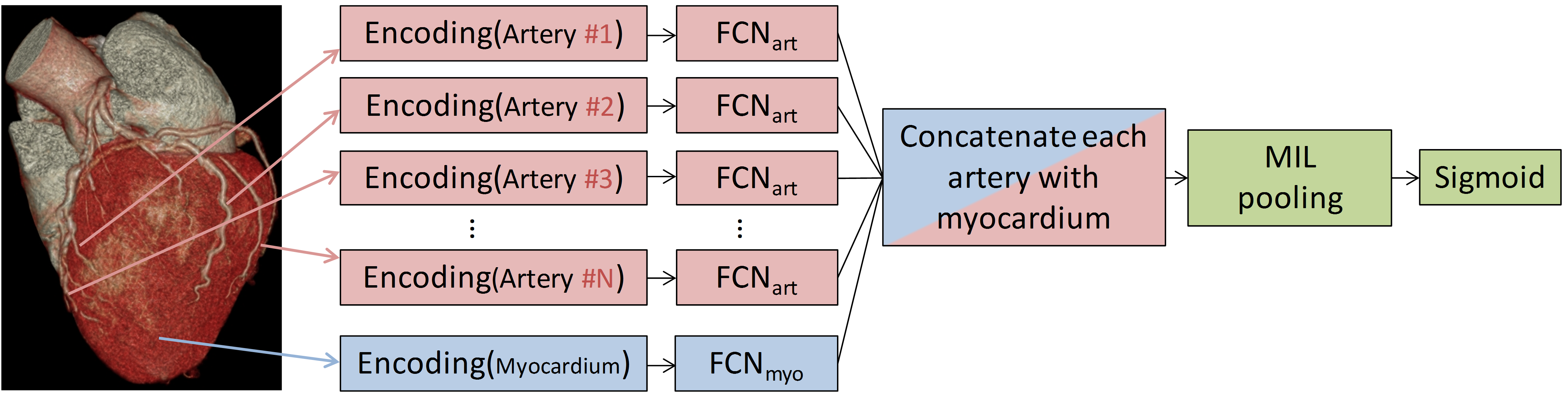}
	\caption{Illustration of the proposed multiple instance learning (MIL) approach.  
		Encodings of N coronary arteries are extracted using the pre-trained CAEs, presented in \cite{zreik2019deep}, and encodings of the LV myocardium are extracted by the pre-trained CAE, presented in \cite{zreik2018deep}. Subsequently, encodings of each artery (out of N extracted arteries) are separately fed to a fully connected neural network ($FCN_{art}$) consisting of 3 fully connected layers with 64 neurons each. The weights of the N $FCN_{art}$ networks are shared. An identical non-shared network ($FCN_{myo}$) is used to analyze the LV myocardium encodings. The outputs of both networks are concatenated to form a bag-of-instances, where each instance contains a low-dimensional embeddings of a single artery concatenated with the low-dimensional embeddings of the LV myocardium. Therefore, each patient is represented by N instances, corresponding to N extracted arteries, which are then fed to a MIL pooling operator \cite{ilse2018attention}. The MIL operator, based on the attention mechanism, uses a trainable weighted average of the instances where the weights are determined by a fully connected 32 neuron neural network. To be invariant to the size of the bag (N; number of arteries), the weights are forced to sum to 1, by using a softmax operator on the output. Finally, the output of the MIL pooling operator is fed into a sigmoid to output a probability to belong to the positive or the negative class. 
	}
	\label{fig:mil_flow}
	
\end{figure*}

Blood flow to the LV myocardium may be affected by one or multiple coronary artery stenoses in one or multiple coronary arteries \cite{Pijl96,koo2011optimal,Tayl13}.
Therefore, local analysis of a single stenosis in a single artery may be insufficient, but the entire coronary artery tree must be taken into account. Moreover, given that functional obstruction of the blood flow in the coronaries may cause ischemia in the LV myocardium, analysis of the myocardium may provide additional valuable information \cite{zreik2018deep,Han17}. 
In clinical practice, only the minimal single FFR value per diseased coronary artery is typically recorded, resulting in a single reference label per artery.
Hence, employing supervised end-to-end machine learning methods (e.g. a 3D-CNN or a recurrent CNN \cite{zreik2018recurrent}) to directly analyze all arteries in the coronary tree together with the LV myocardium to detect the functional significance of a stenosis is infeasible. 
Therefore, in this work, all arteries in the coronary tree and the LV myocardium are separately analyzed by unsupervised CAEs and represented with a fixed number of encodings \cite{zreik2018deep,zreik2019deep}. Then, these encodings are jointly used to classify patients according to the presence of functionally significant stenosis, as determined by the invasive FFR measurement. To handle the varying number of coronary arteries in each patient, the classification is performed using an attention-based multiple instance neural network \cite{ilse2018attention}. An illustration of the proposed method is shown in \cref{fig:mil_flow}.

\subsection{Encoding the artery}

\begin{figure}
	\centering
	\includegraphics[width=1\linewidth]{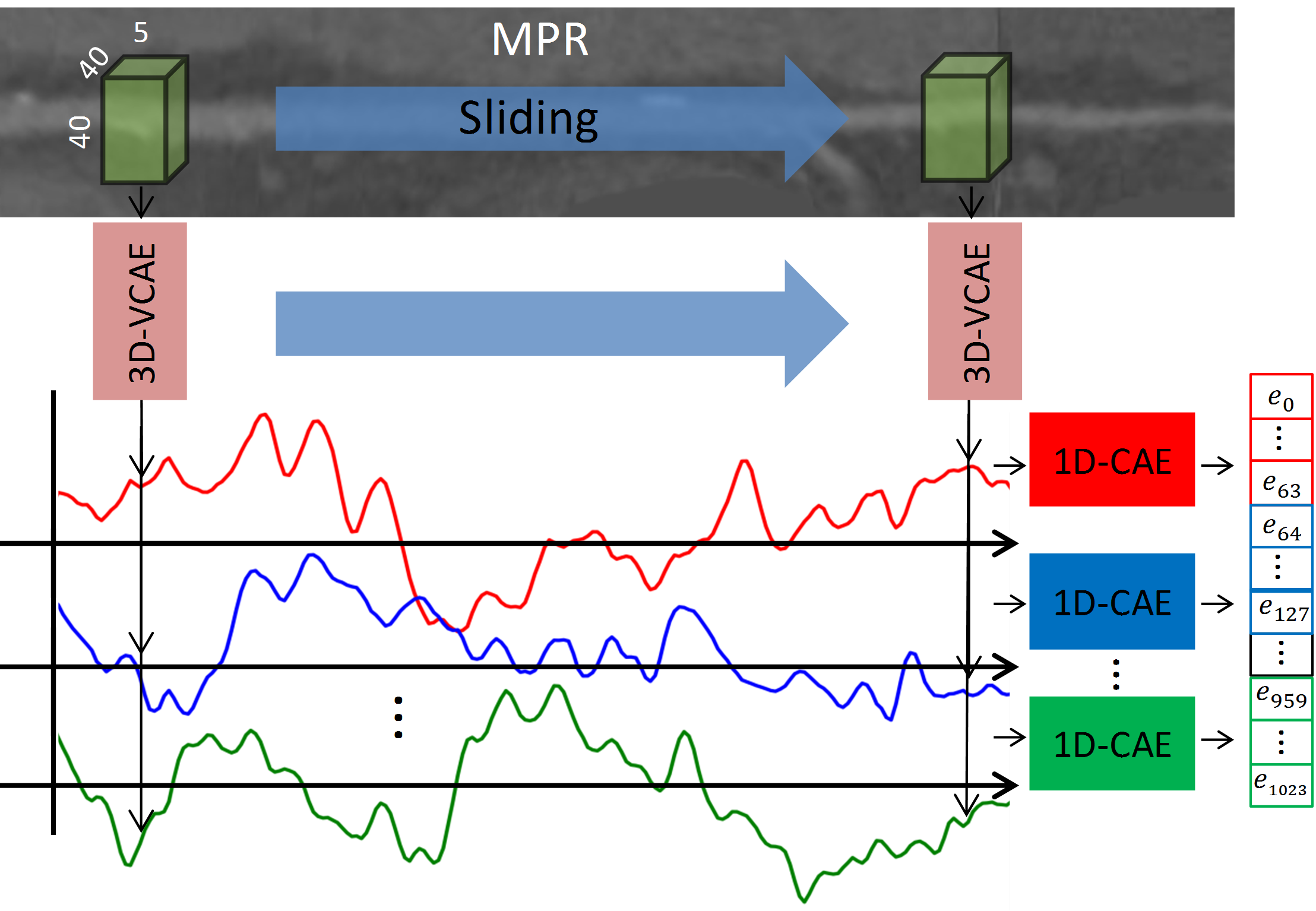}
	\caption{Illustration of the artery encoding. To encode an MPR volume of a complete artery into a fixed number of encodings, a two stage encoding approach is applied. First, a 3D variational convolutional autoencoder (3D-VCAE) is applied to local 40x40x5 voxel sub-volumes extracted from the MPR along the artery. The 3D-VCAE encodes each volume into an encoding in a latent space of 16 dimensions. When applied to all sequential sub-volumes along the artery, the result is a features map of the same height as the number of encodings and the same length as the artery length (L). This features map is then represented as a set of individual 1D sequences of encodings. Each sequence contains an individual value of the encoding in latent space of 16 dimension, running along the artery. Then, a 1D convolutional autoencoder (1D-CAE) is applied separately to each of the 16 sequences of encodings that encodes each sequence further into a second latent space with 64 dimensions. This results in a fixed number of encodings (1024) per artery, that represent the complete artery volume, regardless of its length and shape. 
	}
	\label{fig:encoding_flow_art_myo}
	
\end{figure}

Coronary arteries are complex anatomical 3D structures with varying lengths and anatomies across patients \cite{pannu2003current}. An MPR volume of a single artery contains a large number of voxels (millions). Therefore, to efficiently encode and accurately reconstruct a whole MPR volume of a complete coronary artery, a single CAE may not be used. Therefore, in our previous work \cite{zreik2019deep}, we have employed two disjoint CAEs to encode the complete MPR volume of a single coronary artery. \cref{fig:encoding_flow_art_myo} illustrates the employed encoding workflow. First, a 3D variational convolutional autoencoder (3D-VCAE) is applied to 40x40x5 voxel local sub-volumes extracted from the MPR along the artery centerline. The 3D-VCAE encodes each sub-volume into an encoding of 16 values. When applied sequentially to all sub-volumes along the artery (of length L), the result is a feature map of size 16xL. This feature map is then represented as 16 individual 1D sequences of encodings. Each 1xL sequence contains an individual encoding out of the set of encodings, running along the artery (colored signals in \cref{fig:encoding_flow_art_myo}). To encode arteries with different lengths, sequences of encodings of short arteries were padded into a maximum length 800, which corresponds to the longest artery in the dataset. Then, a 1D-CAE encodes each one of the 16 sequences into a second latent space of 64 dimensions. This results in 1024 (16x64) features representing the complete artery. The architectures and the training details of the 3D-VCAE and 1D-CAE used in this work are described in \cite{zreik2019deep}.

\subsection{Encoding the myocardium}

Given that functionally significant coronary artery stenosis causes ischemia in the LV myocardium and thereby likely changes its texture characteristics in a CCTA image \cite{Niko06,Osaw16,zreik2018deep,Han17}, the LV myocardium is analyzed. First, the LV myocardium is segmented using a multiscale CNN that analyzes two scales (low and high resolution) of three orthogonal (axial, coronal and sagittal) image patches. Characteristics of the segmented LV myocardium and its immediate vicinity are extracted using a CAE, as described in our previous work \cite{zreik2018deep}.
Since a functionally significant stenosis is expected to have a local impact on the myocardial blood perfusion \cite{Mehr11,Ross14}, the LV myocardium is divided into 500 spatially connected clusters. To describe the whole LV myocardium, rather than its clusters, simple statistics of the encodings across the clusters are computed and used as LV myocardium features. As a result, the whole LV myocardium is characterized with a total of 512 features. Full details of the multiscale CNN and CAE architectures and the training details can be found in \cite{zreik2018deep}.

\subsection{Classifying patients according to the presence of functionally significant stenosis}

Based on the extracted encodings of the coronary arteries and the LV myocardium, patients are classified according to the presence of functionally significant stenosis. This was defined by the invasively measured FFR. The clinically common 0.8 threshold on measured FFR was applied \cite{Petr13b} to distinguish between the positive class comprised patients with $FFR\le0.8$, i.e. those having functionally significant stenosis, and the negative class comprised patients with $FFR>0.8$, i.e. those without a functionally significant stenosis.

The classification is performed following a multiple instance learning (MIL) approach \cite{ilse2018attention} and illustrated in \cref{fig:mil_flow}. Functional significance of a stenosis within a specific coronary artery is dependent on the blood flow in the entire coronary tree, consisting of all arteries and their branches \cite{Tayl13}. Therefore, to determine the functional significance of a stenosis in a patient, all arteries in the coronary artery tree are analyzed simultaneously. In addition, as previous studies have shown \cite{zreik2018deep,Han17,van2018deep}, valuable information regarding the significance of a stenosis is present in the LV myocardium. Therefore, in this work, encodings of all coronary arteries are combined with the encodings of the LV myocardium. Subsequently, these combined encodings are used to classify patients, according to the presence of a functionally significant stenosis in one or more coronary arteries. 
To process the varying number of arteries extracted in a CCTA scan, encodings of each of the N arteries are separately fed to a fully connected neural network ($FCN_{art}$ in \cref{fig:mil_flow}) consisting of 3 fully connected layers with 64 neurons each. To limit the number of trainable parameters and reduce the risk of overfitting, the weights of the N $FCN_{art}$ networks are shared, making it practically a single network. An identical non-shared network ($FCN_{myo}$ in \cref{fig:mil_flow}) is used to analyze the LV myocardium encodings. The outputs of all networks are concatenated to form a bag-of-instances; each instance contains a low-dimensional embeddings of a single artery (output of each $FCN_{art}$) concatenated with the low-dimensional embeddings of the LV myocardium (output of $FCN_{myo}$). Hence, each patient is represented by N instances, corresponding to N extracted arteries, which are fed to a MIL pooling operator \cite{ilse2018attention}. The MIL operator, based on the attention mechanism \cite{bahdanau2014neural}, uses a trainable weighted average of the instances, where the weights are determined by a fully connected 32 neuron neural network. To be invariant to the size of the bag (N; number of arteries), the weights are forced to sum to 1, by using a softmax operator on the output. Finally, the output of the MIL pooling operator is fed into a sigmoid to output a probability to belong to the positive or the negative class. Parametric rectified linear units (PReLU) \cite{he2015delving} are used in all layers but the last one. To battle overfitting, a dropout of 0.5 is applied between all fully connected layers in $FCN_{art}$ and $FCN_{myo}$, and L2 regularization is used with $\gamma=0.001$ for all layers.

Performance of patient classification is evaluated using a receiver operating characteristic (ROC) curve, the corresponding area under the ROC curve (AUC), and the sensitivity and specificity metrics.

\section{Experiments and results} \label{results_art_myo}
\subsection{Encoding the coronary arteries and the LV myocardium}\label{encoding_art_myo}
To encode the coronary arteries and the LV myocardium in the set of 126 CCTA scans, the pre-trained CNN and CAEs that were previously trained in \cite{zreik2019deep} and \cite{zreik2018deep} were used. Therefore, the training processes of the CNN and the CAEs are not described here and can be found in detail in the original manuscripts. However, to present a valid evaluation of the performance, none of the CCTA scans previously used to train the CNN and CAEs were part of the 126 patients used in this study. 

\begin{figure*}
	\centering
	\includegraphics[width=1.0\textwidth]{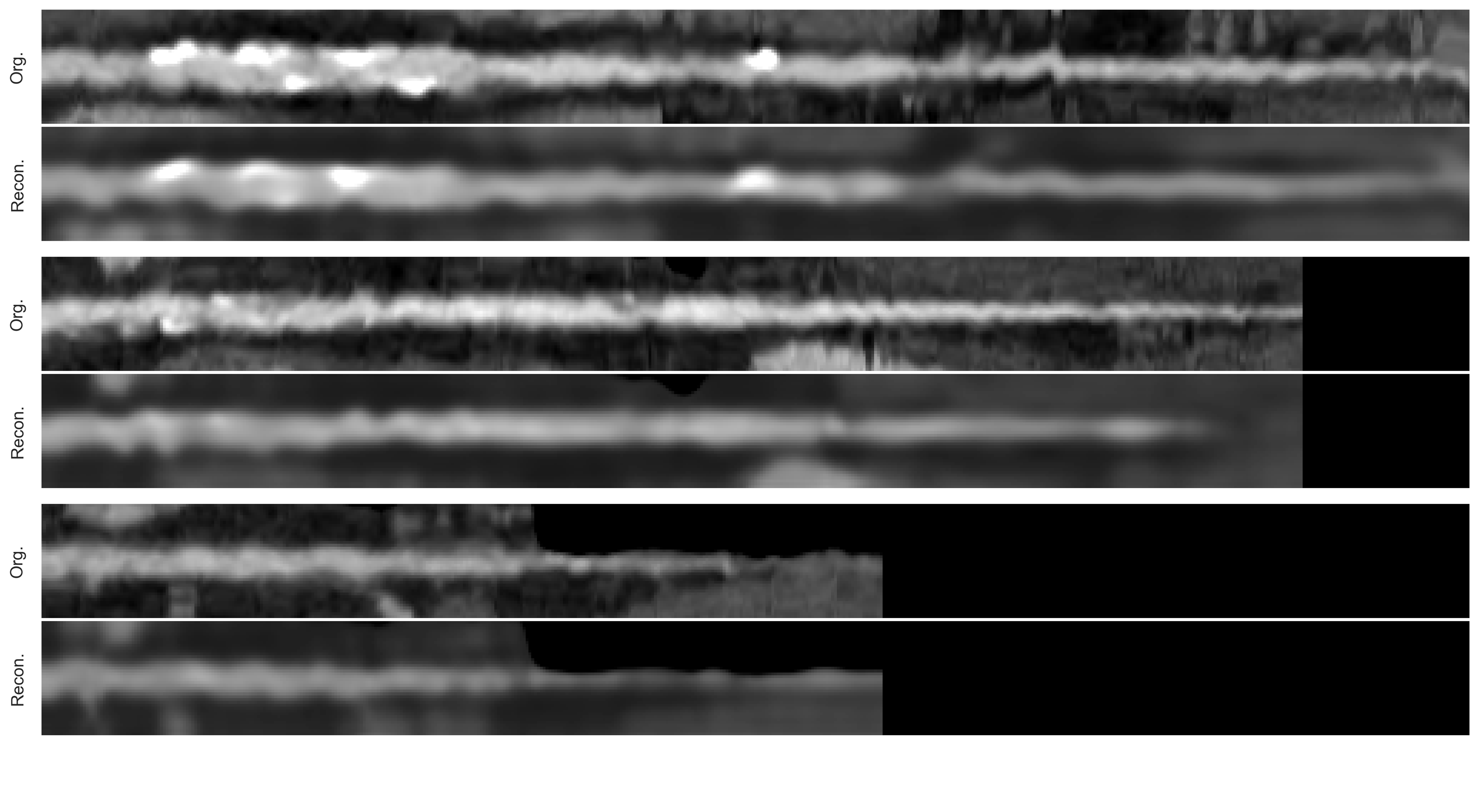}
	\caption{Examples of three MPR volumes of complete coronary arteries (\textit{Org.}), that were encoded and reconstructed (\textit{Recon.}), using the pre-trained disjoint CAEs as presented in \cite{zreik2019deep}. Arteries with three different lengths and levels of classifications are presented.}
	\label{fig:all_reconstructions_art_myo}
	
\end{figure*}

To extract encodings from the coronary arteries, the pre-trained 3D-VCAE and 1D-CAE described in \cite{zreik2019deep} were applied to all extracted arteries in all 126 CCTA scans. To illustrate that the extracted encodings contain enough information to retrieve and reconstruct the input volume, \cref{fig:all_reconstructions_art_myo} shows three randomly selected examples of encoded and reconstructed MPR volumes of complete arteries.

\begin{figure}
	\centering
	\includegraphics[width=1.0\textwidth]{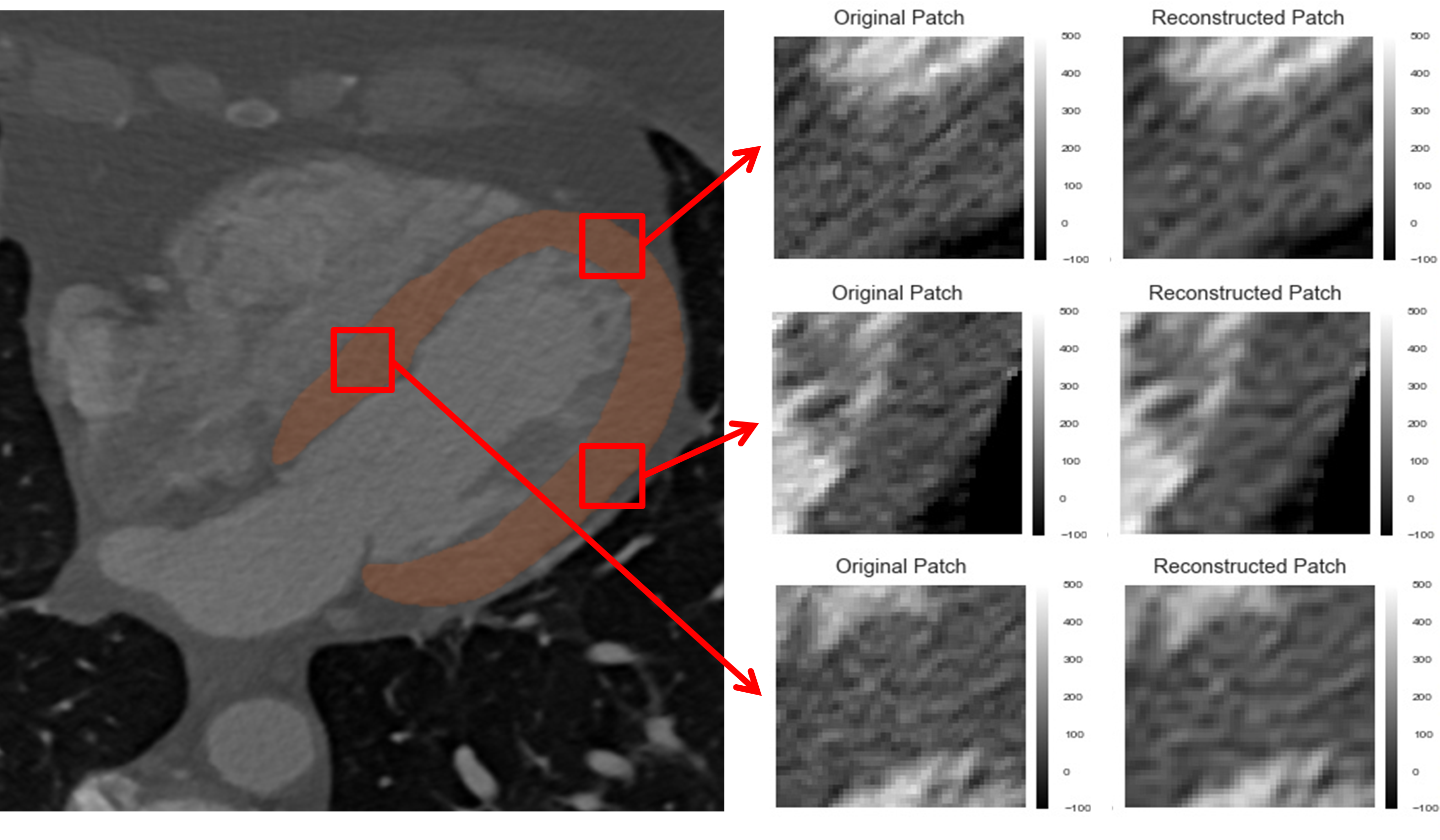}
	\caption{Left: An example of an axial view of an CCTA scan, with the automatically segmented LV myocardium (red overlay) using the pre-trained multiscale CNN. Right: Within the segmented LV myocardium, 3 randomly selected myocardial image patches (\textit{Original}) are encoded and then reconstructed (\textit{Reconstructed}) by the pre-trained CAE, as presented in \cite{zreik2018deep}.}
	\label{fig:myo_enc}
	
\end{figure}

To extract encodings and features of the LV myocardium, the pre-trained multiscale CNN and CAE described in \cite{zreik2018deep} were employed in all 126 CCTA scans to segment and encode the LV myocardium, respectively. \cref{fig:myo_enc} illustrates a segmented LV myocardium and three randomly selected LV myocardium axial patches that were encoded and thereafter reconstructed back. 

\subsection{Classifying patients according to the presence of functionally significant stenosis}

Classification of patients was performed using the extracted encodings of the coronary arteries and the LV myocardium.
To assess the classification performance and its robustness, stratified 5-fold cross-validation experiments were performed. In each experiment, all arteries and the LV myocardium of 101 (80\%) patients were used to train the classification network (\cref{fig:mil_flow}), while the remaining 25 (20\%) patients were left out to test the performance. 

\begin{figure}
	\centering
	\includegraphics[width=1\textwidth]{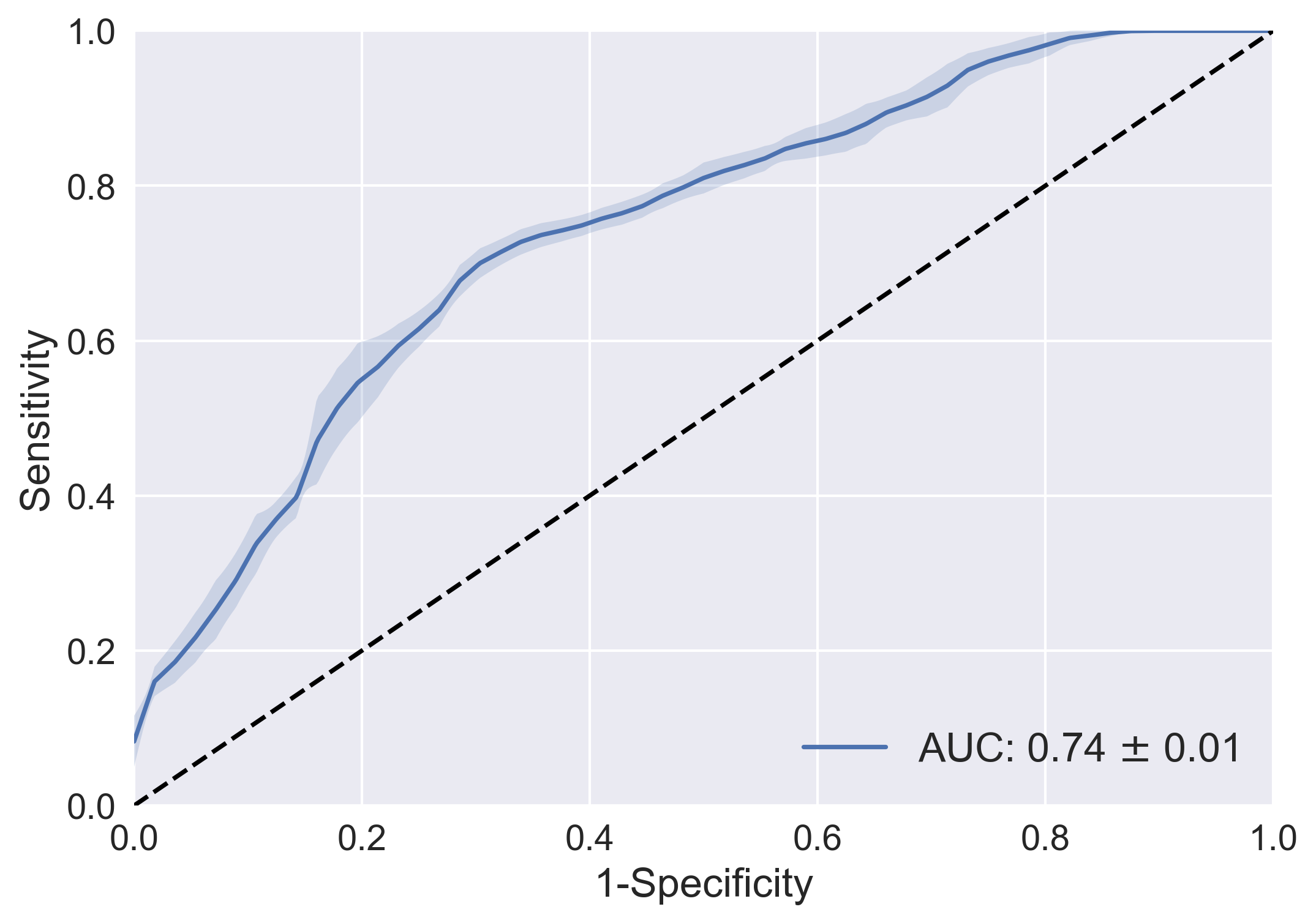}
	\caption{Average ROC curve, and corresponding area under curve (AUC), for classification of patients according to FFR. The classification was performed employing the extracted encodings of the arteries and the LV myocardium using an attention-based MIL neural network. The shaded area represent the standard deviation of the sensitivity.}
	\label{fig:cv_roc_art_myo}
	
\end{figure}

As the number of arteries varies between patients, training the classification network was performed using mini-batches of 1 patient to minimize the binary cross entropy loss function using the sigmoid output. In each mini-batch, encodings of all coronary arteries (N) and the LV myocardium of the same patient were included.
Training was performed for 200,000 iterations, where every 1000 iterations, the latest model was saved. To assess the variability of the performance, in the end of the training, the last 10 saved models were used to evaluate the performance on the test set. The obtained results are shown in \cref{fig:cv_roc_art_myo}. An average AUC of $0.74\pm0.01$ was achieved. At the sensitivity of 0.70, the achieved specificity was 0.70. \cref{tab:perf_ffr_art_myo} shows the obtained average sensitivities and specificities in three different ranges of FFR values.

\begin{table}
	\centering
	\begin{tabular}{@{}lcc@{}}
		\toprule
		& Sensitivity & Specificity \\ \midrule
		$FFR\leq0.7 $ & 0.82        & -           \\ 
		$0.7<FFR<0.9$ & 0.63        & 0.70        \\ 
		$FFR\geq0.9$  & -           & 0.70        \\ \bottomrule
	\end{tabular}
	\caption{Average sensitivity and specificity for detection of patients with functionally significant coronary artery stenosis in three different ranges of FFR measurements.}
	\label{tab:perf_ffr_art_myo}
\end{table}

\begin{figure}
	\centering
	\includegraphics[width=1\textwidth]{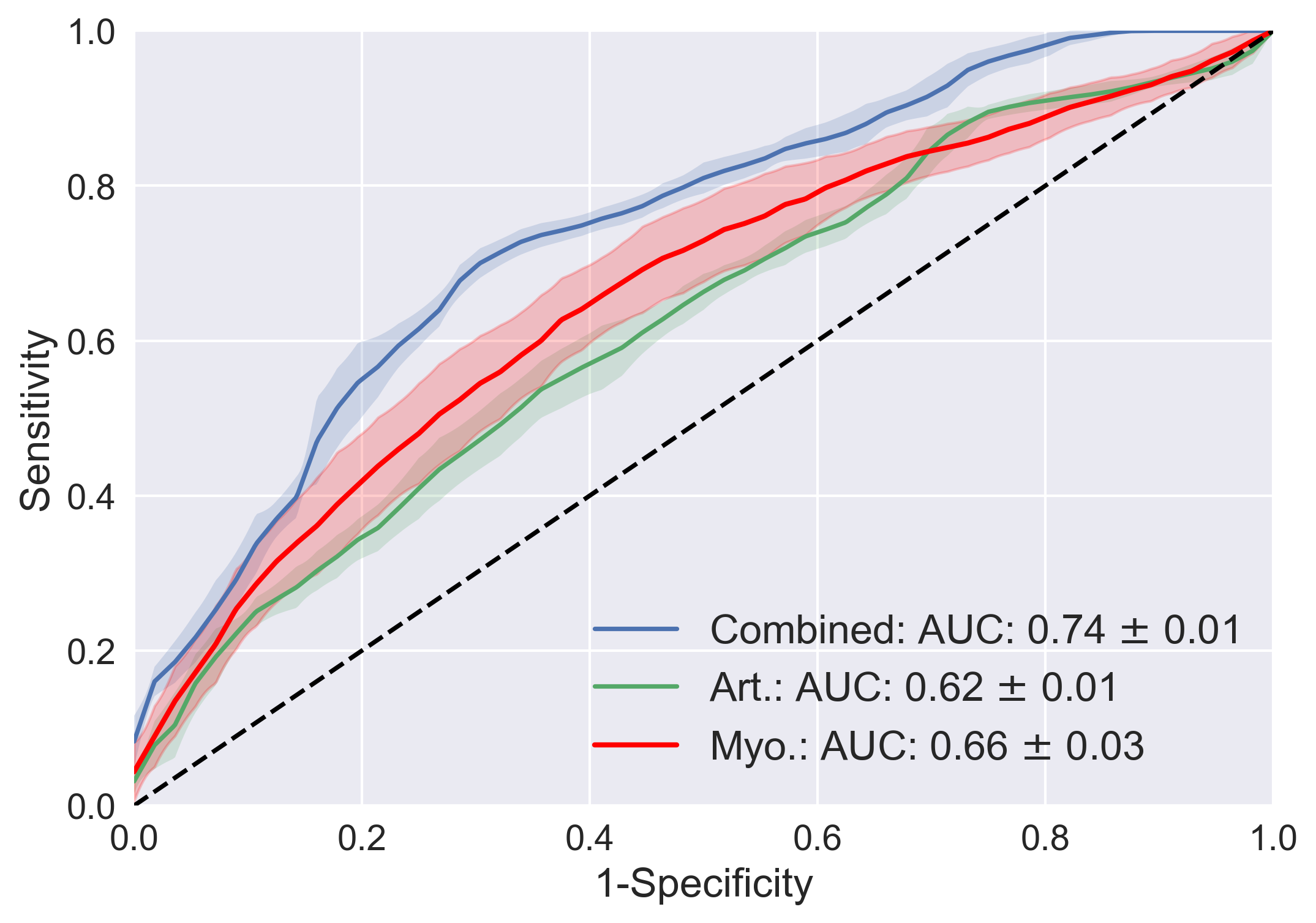}
	\caption{Average ROC curves, and corresponding area under curve (AUC), for classification of patients according to FFR using (a) encodings of the arteries only (\textit{Art.}), (b) encodings of the LV myocardium only (\textit{Myo.}) and (c) encodings of both the LV myocardium and the coronary arteries (\textit{Combined}). The classification was performed employing the extracted encodings using an attention-based MIL neural network. The shaded areas represent the standard deviation of the sensitivity across the 10 evaluated models.}
	\label{fig:cv_combined_art_myo}
	
\end{figure}

To assess the importance of combining the information extracted from the LV myocardium on the classification performance, an additional experiment was performed. In this experiment, only the encodings of the coronary arteries were used, without incorporating the LV myocardium encodings. To do so, an identical classification network was employed, while omitting the LV myocardium subnetwork ($FCN_{myo}$ in \cref{fig:mil_flow}). Training this classification network was performed in the same manner as the full network. The achieved performance, compared to the performance of the combined analysis and to the LV myocardium only analysis (as reported in \cite{zreik2018deep}) is shown in \cref{fig:cv_combined_art_myo} and in \cref{list:compare_art_myo}.

\begin{table}
	\centering
	\caption{Area under ROC curve (AUC), sensitivity and specificity achieved for classification with different combinations of encodings; \textit{Arteries only} where the LV myocardium encodings were not incorporated, \textit{Myocardium only} where the arteries encodings were not analyzed, and the proposed \textit{Combined} analysis. }
	\label{list:compare_art_myo}
	\begin{tabular}{lccc}
		\toprule
		\multicolumn{1}{c}{} & AUC & Sensitivity & Specificity  \\ \midrule
		\multicolumn{1}{l}{Arteries only} & $0.62\pm0.01$ & 0.70 & 0.46  \\  
		\multicolumn{1}{l}{Myocardium only} & $0.66\pm0.03$ & 0.70 & 0.55  \\  
		\multicolumn{1}{l}{\textbf{Combined}} & \textbf{0.74$\pm$0.01} & \textbf{0.70} & \textbf{0.70} \\ \bottomrule
	\end{tabular}
\end{table}

\subsection{Comparison with other FFR classification methods}\label{others_art_myo}

We compare the classification results of the here presented method with the results of previous methods. These methods either analyzed of the blood flow in the coronary arteries or analyzed the LV myocardium. \cref{list:other_methods_results_art_myo} lists the results as reported in the original works. 
The table demonstrates that the methods achieving highest accuracy and AUC relied on blood flow analysis in the coronary arteries. Moreover, the here presented method seems to outperform the methods that analyze only on the LV myocardium. However, different methods were evaluated with different datasets, using different patients cohorts. Therefore, this comparison only indicates the differences in performance and should be considered with caution.

\begin{table}
		\resizebox{\textwidth}{!}{%
	\centering
	\caption{Performance comparison with previous work. Table lists number of evaluated patients and arteries and the achieved patient-level accuracy and the area under the ROC curve (AUC) for detection of functionally significant stenosis as reported in the original studies. Please note that these methods perform different analyses: either analyzing of the blood flow in the coronary arteries (\textit{Flow}), detecting ischemic changes directly in LV myocardium (\textit{Myo.}), or, as proposed in this work; combining the analysis of coronary arteries with the LV myocardium. }
	\label{list:other_methods_results_art_myo}

	\begin{tabular}{cccccc}
		\toprule

		& \textbf{Study}       & \multicolumn{1}{l}{\textbf{Patients}} & \multicolumn{1}{l}{\textbf{arteries}} & \multicolumn{1}{l}{\textbf{Accuracy}} & \multicolumn{1}{l}{\textbf{AUC}} \\ \midrule
		\multirow{2}{*}{\rotatebox[origin=c]{90}{Flow}} 
		& N{\o}rgaard et al.\cite{Norg14}        & 254                                       & 484                                      & 0.81                                  & 0.90                                                          \\ 
		& Coenen et al.\cite{coenen2018diagnostic}        & 303                                       & 525                                      & 0.71                                     & -                                                                
		\\ \midrule
		
		\multirow{2}{*}{\rotatebox[origin=c]{90}{Myo.}} 
		
		& Zreik et al.\cite{zreik2018deep}                 & 126                                       & -                                        & 0.63                                  & 0.66                              
		\\ 
		
		& Han et al.\cite{Han17}         & 252                                       & 407                                      & 0.63                                  & -

		\\ \midrule
		
		\multirow{2}{*}{\rotatebox[origin=c]{90}{}} & Proposed         & 126                                       & -                                      & 0.70                                  & 0.74

		\\ \bottomrule	                
	\end{tabular}%
}
\end{table}

\section{Discussion}\label{discussion_art_myo}

A method for automatic detection of patients with functionally significant stenosis in the coronary arteries in CCTA scans has been presented. 
First, our previous work is employed to characterize the coronary arteries \cite{zreik2019deep} and the LV myocardium \cite{zreik2018deep} in an unsupervised fashion. These extracted characteristics are subsequently used to classify patients with a supervised MIL operator, employing an attention-based trainable neural network \cite{ilse2018attention}. The results demonstrate that the proposed combined analysis outperforms the analysis of the coronary arteries or LV myocardium only. To the best of our knowledge, this is the first method to combine information derived from the coronary arteries as well as the LV myocardium.

Given the high resolution of modern CT scanners, the volumes of the complete coronary arteries and the LV myocardium are large. Moreover, the reference FFR is reported only on the artery level. Therefore, directly detecting the functional significance of a stenosis (e.g. with supervised 3D-CNN) is far from trivial.
Hence, in this work, we have used unsupervised CAEs to characterize and encode both the arteries and the LV myocardium before employing a supervised neural network. Given the limited number of patients included in this study, training the system end-to-end, with its three components, is hardly feasible due to overfitting. Given a larger dataset, training all components end-to-end could potentially make the extracted encodings more discriminative with respect to the classification task, and therefore further improve the performance. Future work might investigate this.

Moderate performance was achieved for detection of patients with functionally significant stenosis, while relying only on unsupervised features extraction methods for both the coronary arteries and the LV myocardium in a CCTA (\cref{fig:cv_roc_art_myo}). This combined analysis substantially improves the low specificity of visual inspection of CCTA only ($<50\%$) \cite{Ko12} and could potentially reduce the number of unnecessary ICA procedures in patients with suspected obstructive CAD. Moreover, our previous work \cite{van2018deep}, presented a combination of expert reading and automatic analysis of the LV myocardoum only and showed an improvement in diagnostic accuracy in patients with intermediate degree of stenosis. Similarly, future work might evaluate whether the proposed combined analysis is of additional value to the expert reading in the clinical settings.

The proposed method has higher diagnostic accuracy than methods that analyze only the LV myocardium, but lower than methods that analyze blood flow in the coronary arteries. However, flow-based methods heavily depend on accurate lumen segmentation, which is a very cumbersome task that typically requires manual interaction of an expert to correct the automatically segmented lumen, especially in CCTA scans with excessive atherosclerotic calcifications \cite{Kiri13a}.
As a result of this requirement, heavily diseased patients with large calcifications or coronary stents, or CCTA images with imaging artifacts are typically not eligible for such analyses\cite{lu2017noninvasive,Pontone2019,Sakuma2019}.
In contrast, instead of artery lumen segmentation, the here proposed method requires only the coronary artery centerline as an input. Extraction of the artery centerline is to some extent a simple task as compared to the artery lumen segmentation. In this work, we have employed our previous work to extract the coronary artery centerlines \cite{wolterink2019coronary}, however, any other available extraction method could be used instead.

The contribution of the combined analysis was investigated. \cref{list:compare_art_myo} and \cref{fig:cv_combined_art_myo} demonstrate that the proposed combined analysis outperforms the analysis of the coronary arteries or the LV myocardium only. This result is on par with our assumptions that, in respect to the functional significance of a coronary stenosis, a combination of the ischemic changes present in the LV myocardium \cite{zreik2018deep}, and in the anatomical characteristics of the coronary arteries \cite{zreik2019deep}, is beneficial for identifying these patients. However, in the presented method, features of the whole LV myocardium were concatenated to the encoding of each coronary artery without any alignment to blood diffusion territories. As a myocardial territory perfused by a stenotic coronary artery might have local ischemia \cite{Mehr11,Ross14}, analyzing the LV myocardium as a whole might have masked these local subtle changes. Although this ad-hoc concatenation was proven beneficial (\cref{list:compare_art_myo} and \cref{fig:cv_combined_art_myo}), concatenating each artery with the corresponding local perfusion region may further improve the performance. To perform such local concatenation, the segmented LV myocardium could be aligned with the standard 17-segments model \cite{Cerq02a} or with a personalized perfusion model \cite{jung2016patient}. Thereafter, encodings of each LV myocardium segment could be concatenated to the corresponding coronary artery encoding. Future work might address this.

To conclude, this study presented an automatic method for detection of patients with functionally significant stenosis in the coronary arteries in CCTA images. 
The method is based on two unsupervised methods that separately encode the coronary arteries and the LV myocardium. Thereafter, a supervised MIL neural network combines the two sets of encodings and classifies patients according to the presence of functionally significant stenosis. The achieved moderate classification performance shows the feasibility of reducing the number of patients that unnecessarily undergo invasive FFR measurements.
	
\section*{Acknowledgments}

This study was financially supported by the project FSCAD, funded by the Netherlands Organization for Health Research and Development (ZonMw) in the framework of the research programme IMDI (Innovative Medical Devices Initiative); project 104003009. We gratefully acknowledge the support of NVIDIA Corporation with the donation of the Tesla K40 GPU used for this research.

\bibliographystyle{IEEEtran}

\begin{thebibliography}{10}
	\providecommand{\url}[1]{#1}
	\csname url@samestyle\endcsname
	\providecommand{\newblock}{\relax}
	\providecommand{\bibinfo}[2]{#2}
	\providecommand{\BIBentrySTDinterwordspacing}{\spaceskip=0pt\relax}
	\providecommand{\BIBentryALTinterwordstretchfactor}{4}
	\providecommand{\BIBentryALTinterwordspacing}{\spaceskip=\fontdimen2\font plus
		\BIBentryALTinterwordstretchfactor\fontdimen3\font minus
		\fontdimen4\font\relax}
	\providecommand{\BIBforeignlanguage}[2]{{%
			\expandafter\ifx\csname l@#1\endcsname\relax
			\typeout{** WARNING: IEEEtran.bst: No hyphenation pattern has been}%
			\typeout{** loaded for the language `#1'. Using the pattern for}%
			\typeout{** the default language instead.}%
			\else
			\language=\csname l@#1\endcsname
			\fi
			#2}}
	\providecommand{\BIBdecl}{\relax}
	\BIBdecl
	
	\bibitem{benjamin2018heart}
	E.~J. Benjamin, S.~S. Virani, C.~W. Callaway, A.~M. Chamberlain, A.~R. Chang,
	S.~Cheng, S.~E. Chiuve, M.~Cushman, F.~N. Delling, R.~Deo \emph{et~al.},
	``Heart disease and stroke statistics-2018 update: a report from the american
	heart association.'' \emph{Circulation}, vol. 137, no.~12, p. e67, 2018.
	
	\bibitem{cury2016cad}
	R.~C. Cury, S.~Abbara, S.~Achenbach, A.~Agatston, D.~S. Berman, M.~J. Budoff,
	K.~E. Dill, J.~E. Jacobs, C.~D. Maroules, G.~D. Rubin \emph{et~al.},
	``{CAD-RADSTM} coronary artery disease--reporting and data system. an expert
	consensus document of the society of cardiovascular computed tomography
	({SCCT}), the american college of radiology ({ACR}) and the north american
	society for cardiovascular imaging ({NASCI}). endorsed by the american
	college of cardiology,'' \emph{Journal of Cardiovascular Computed
		Tomography}, vol.~10, no.~4, pp. 269--281, 2016.
	
	\bibitem{Pijl96}
	N.~H. Pijls, B.~de~Bruyne, K.~Peels, P.~H. van~der Voort, H.~J. Bonnier,
	J.~Bartunek, and J.~J. Koolen, ``Measurement of fractional flow reserve to
	assess the functional severity of coronary-artery stenoses,'' \emph{New
		England Journal of Medicine}, vol. 334, no.~26, pp. 1703--1708, 1996.
	
	\bibitem{Toni09}
	P.~A. Tonino, B.~De~Bruyne, N.~H. Pijls, U.~Siebert, F.~Ikeno, M.~vant Veer,
	V.~Klauss, G.~Manoharan, T.~Engstr{\o}m, K.~G. Oldroyd \emph{et~al.},
	``Fractional flow reserve versus angiography for guiding percutaneous
	coronary intervention,'' \emph{New England Journal of Medicine}, vol. 360,
	no.~3, pp. 213--224, 2009.
	
	\bibitem{Pijl10}
	N.~H. Pijls, W.~F. Fearon, P.~A. Tonino, U.~Siebert, F.~Ikeno, B.~Bornschein,
	M.~van't Veer, V.~Klauss, G.~Manoharan, T.~Engstr{\o}m \emph{et~al.},
	``Fractional flow reserve versus angiography for guiding percutaneous
	coronary intervention in patients with multivessel coronary artery disease:
	2-year follow-up of the {FAME} (fractional flow reserve versus angiography
	for multivessel evaluation) study,'' \emph{Journal of the American College of
		Cardiology}, vol.~56, no.~3, pp. 177--184, 2010.
	
	\bibitem{Nune15}
	L.~X. van Nunen, F.~M. Zimmermann, P.~A. Tonino, E.~Barbato, A.~Baumbach,
	T.~Engstr{\o}m, V.~Klauss, P.~A. MacCarthy, G.~Manoharan, K.~G. Oldroyd
	\emph{et~al.}, ``Fractional flow reserve versus angiography for guidance of
	{PCI} in patients with multivessel coronary artery disease ({FAME}): 5-year
	follow-up of a randomised controlled trial,'' \emph{The Lancet}, vol. 386,
	no. 10006, pp. 1853--1860, 2015.
	
	\bibitem{Budo08a}
	M.~J. Budoff, D.~Dowe, J.~G. Jollis, M.~Gitter, J.~Sutherland, E.~Halamert,
	M.~Scherer, R.~Bellinger, A.~Martin, R.~Benton \emph{et~al.}, ``Diagnostic
	performance of 64-multidetector row coronary computed tomographic angiography
	for evaluation of coronary artery stenosis in individuals without known
	coronary artery disease: results from the prospective multicenter {ACCURACY}
	trial,'' \emph{Journal of the American College of Cardiology}, vol.~52,
	no.~21, pp. 1724--1732, 2008.
	
	\bibitem{Meij08}
	W.~B. Meijboom, C.~A. Van~Mieghem, N.~van Pelt, A.~Weustink, F.~Pugliese, N.~R.
	Mollet, E.~Boersma, E.~Regar, R.~J. van Geuns, P.~J. de~Jaegere
	\emph{et~al.}, ``Comprehensive assessment of coronary artery stenoses:
	computed tomography coronary angiography versus conventional coronary
	angiography and correlation with fractional flow reserve in patients with
	stable angina,'' \emph{Journal of the American College of Cardiology},
	vol.~52, no.~8, pp. 636--643, 2008.
	
	\bibitem{Bamb11}
	F.~Bamberg, A.~Becker, F.~Schwarz, R.~P. Marcus, M.~Greif, F.~von Ziegler,
	R.~Blankstein, U.~Hoffmann, W.~H. Sommer, V.~S. Hoffmann \emph{et~al.},
	``Detection of hemodynamically significant coronary artery stenosis:
	incremental diagnostic value of dynamic {CT}-based myocardial perfusion
	imaging,'' \emph{Radiology}, vol. 260, no.~3, pp. 689--698, 2011.
	
	\bibitem{Ko12}
	B.~S. Ko, J.~D. Cameron, M.~Leung, I.~T. Meredith, D.~P. Leong, P.~R. Antonis,
	M.~Crossett, J.~Troupis, R.~Harper, Y.~Malaiapan \emph{et~al.}, ``Combined
	{CT} coronary angiography and stress myocardial perfusion imaging for
	hemodynamically significant stenoses in patients with suspected coronary
	artery disease: a comparison with fractional flow reserve,'' \emph{JACC:
		Cardiovascular Imaging}, vol.~5, no.~11, pp. 1097--1111, 2012.
	
	\bibitem{Tayl13}
	C.~A. Taylor, T.~A. Fonte, and J.~K. Min, ``Computational fluid dynamics
	applied to cardiac computed tomography for noninvasive quantification of
	fractional flow reserve: Scientific basis,'' \emph{Journal of the American
		College of Cardiology}, vol.~61, no.~22, pp. 2233 -- 2241, 2013.
	
	\bibitem{Itu12}
	L.~Itu, P.~Sharma, V.~Mihalef, A.~Kamen, C.~Suciu, and D.~Lomaniciu, ``A
	patient-specific reduced-order model for coronary circulation,'' in
	\emph{Biomedical Imaging (ISBI), 2012 9th IEEE International Symposium
		on}.\hskip 1em plus 0.5em minus 0.4em\relax IEEE, 2012, pp. 832--835.
	
	\bibitem{Nick15}
	H.~Nickisch, Y.~Lamash, S.~Prevrhal, M.~Freiman, M.~Vembar, L.~Goshen, and
	H.~Schmitt, ``Learning patient-specific lumped models for interactive
	coronary blood flow simulations,'' in \emph{International Conference on
		Medical Image Computing and Computer-Assisted Intervention}.\hskip 1em plus
	0.5em minus 0.4em\relax Springer, 2015, pp. 433--441.
	
	\bibitem{itu2016machine}
	L.~Itu, S.~Rapaka, T.~Passerini, B.~Georgescu, C.~Schwemmer, M.~Schoebinger,
	T.~Flohr, P.~Sharma, and D.~Comaniciu, ``A machine-learning approach for
	computation of fractional flow reserve from coronary computed tomography,''
	\emph{Journal of Applied Physiology}, vol. 121, no.~1, pp. 42--52, 2016.
	
	\bibitem{Norg14}
	B.~L. N{\o}rgaard, J.~Leipsic, S.~Gaur, S.~Seneviratne, B.~S. Ko, H.~Ito, J.~M.
	Jensen, L.~Mauri, B.~De~Bruyne, H.~Bezerra \emph{et~al.}, ``Diagnostic
	performance of noninvasive fractional flow reserve derived from coronary
	computed tomography angiography in suspected coronary artery disease: the
	{NXT} trial (analysis of coronary blood flow using {CT} angiography: Next
	steps),'' \emph{Journal of the American College of Cardiology}, vol.~63,
	no.~12, pp. 1145--1155, 2014.
	
	\bibitem{tesche2017coronary}
	C.~Tesche, C.~N. De~Cecco, M.~H. Albrecht, T.~M. Duguay, R.~R. Bayer, S.~E.
	Litwin, D.~H. Steinberg, and U.~J. Schoepf, ``Coronary ct
	angiography--derived fractional flow reserve,'' \emph{Radiology}, vol. 285,
	no.~1, pp. 17--33, 2017.
	
	\bibitem{Kiri13a}
	H.~Kiri{\c{s}}li, M.~Schaap, C.~Metz, A.~Dharampal, W.~B. Meijboom, S.-L.
	Papadopoulou, A.~Dedic, K.~Nieman, M.~De~Graaf, M.~Meijs \emph{et~al.},
	``Standardized evaluation framework for evaluating coronary artery stenosis
	detection, stenosis quantification and lumen segmentation algorithms in
	computed tomography angiography,'' \emph{Medical Image Analysis}, vol.~17,
	no.~8, pp. 859--876, 2013.
	
	\bibitem{zreik2018deep}
	M.~Zreik, N.~Lessmann, R.~W. van Hamersvelt, J.~M. Wolterink, M.~Voskuil, M.~A.
	Viergever, T.~Leiner, and I.~I{\v{s}}gum, ``Deep learning analysis of the
	myocardium in coronary {CT} angiography for identification of patients with
	functionally significant coronary artery stenosis,'' \emph{Medical image
		analysis}, vol.~44, pp. 72--85, 2018.
	
	\bibitem{van2018deep}
	R.~W. van Hamersvelt, M.~Zreik, M.~Voskuil, M.~A. Viergever, I.~I{\v{s}}gum,
	and T.~Leiner, ``Deep learning analysis of left ventricular myocardium in ct
	angiographic intermediate-degree coronary stenosis improves the diagnostic
	accuracy for identification of functionally significant stenosis,''
	\emph{European radiology}, vol.~29, no.~5, pp. 2350--2359, 2019.
	
	\bibitem{Han17}
	D.~Han, J.~H. Lee, A.~Rizvi, H.~Gransar, L.~Baskaran, J.~Schulman-Marcus,
	B.~{\'o}~Hartaigh, F.~Y. Lin, and J.~K. Min, ``Incremental role of resting
	myocardial computed tomography perfusion for predicting physiologically
	significant coronary artery disease: A machine learning approach,''
	\emph{Journal of Nuclear Cardiology}, pp. 1--11, 2017.
	
	\bibitem{Xion15}
	G.~Xiong, D.~Kola, R.~Heo, K.~Elmore, I.~Cho, and J.~K. Min, ``Myocardial
	perfusion analysis in cardiac computed tomography angiographic images at
	rest,'' \emph{Medical Image Analysis}, vol.~24, no.~1, pp. 77--89, 2015.
	
	\bibitem{Cerq02a}
	M.~D. Cerqueira, N.~J. Weissman, V.~Dilsizian, A.~K. Jacobs, S.~Kaul, W.~K.
	Laskey, D.~J. Pennell, J.~A. Rumberger, T.~Ryan, and M.~S. Verani,
	``Standardized myocardial segmentation and nomenclature for tomographic
	imaging of the heart: a statement for healthcare professionals from the
	cardiac imaging committee of the council on clinical cardiology of the
	american heart association,'' \emph{Circulation}, vol. 105, no.~4, pp.
	539--542, 2002.
	
	\bibitem{zreik2019deep}
	M.~Zreik, R.~W. van Hamersvelt, N.~Khalili, J.~M. Wolterink, M.~Voskuil, M.~A.
	Viergever, T.~Leiner, and I.~I{\v{s}}gum, ``Deep learning analysis of
	coronary arteries in cardiac {CT} angiography for detection of patients
	requiring invasive coronary angiography,'' \emph{IEEE Transactions on Medical
		Imaging}, 2019, in press.
	
	\bibitem{ilse2018attention}
	M.~Ilse, J.~M. Tomczak, and M.~Welling, ``Attention-based deep multiple
	instance learning,'' \emph{arXiv preprint arXiv:1802.04712}, 2018.
	
	\bibitem{wolterink2019coronary}
	J.~M. Wolterink, R.~W. van Hamersvelt, M.~A. Viergever, T.~Leiner, and
	I.~I{\v{s}}gum, ``Coronary artery centerline extraction in cardiac {CT}
	angiography using a {CNN-based} orientation classifier,'' \emph{Medical image
		analysis}, vol.~51, pp. 46--60, 2019.
	
	\bibitem{koo2011optimal}
	B.-K. Koo, H.-M. Yang, J.-H. Doh, H.~Choe, S.-Y. Lee, C.-H. Yoon, Y.-K. Cho,
	C.-W. Nam, S.-H. Hur, H.-S. Lim \emph{et~al.}, ``Optimal intravascular
	ultrasound criteria and their accuracy for defining the functional
	significance of intermediate coronary stenoses of different locations,''
	\emph{JACC: Cardiovascular Interventions}, vol.~4, no.~7, pp. 803--811, 2011.
	
	\bibitem{zreik2018recurrent}
	M.~{Zreik}, R.~W. {van Hamersvelt}, J.~M. {Wolterink}, T.~{Leiner}, M.~A.
	{Viergever}, and I.~{Išgum}, ``A recurrent cnn for automatic detection and
	classification of coronary artery plaque and stenosis in coronary {CT}
	angiography,'' \emph{IEEE Transactions on Medical Imaging}, vol.~38, no.~7,
	pp. 1588--1598, July 2019.
	
	\bibitem{pannu2003current}
	H.~K. Pannu, T.~G. Flohr, F.~M. Corl, and E.~K. Fishman, ``Current concepts in
	multi--detector row {CT} evaluation of the coronary arteries: principles,
	techniques, and anatomy,'' \emph{Radiographics}, vol.~23, no. suppl\_1, pp.
	S111--S125, 2003.
	
	\bibitem{Niko06}
	K.~Nikolaou, A.~Knez, C.~Rist, B.~J. Wintersperger, A.~Leber, T.~Johnson, M.~F.
	Reiser, and C.~R. Becker, ``Accuracy of 64-mdct in the diagnosis of ischemic
	heart disease,'' \emph{American Journal of Roentgenology}, vol. 187, no.~1,
	pp. 111--117, 2006.
	
	\bibitem{Osaw16}
	K.~Osawa, T.~Miyoshi, T.~Miki, Y.~Koyama, S.~Sato, S.~Kanazawa, and H.~Ito,
	``Diagnostic performance of first-pass myocardial perfusion imaging without
	stress with computed tomography (ct) compared with coronary ct angiography
	alone, with fractional flow reserve as the reference standard,'' \emph{PloS
		one}, vol.~11, no.~2, p. e0149170, 2016.
	
	\bibitem{Mehr11}
	V.~C. Mehra, C.~Valdiviezo, A.~Arbab-Zadeh, B.~S. Ko, S.~K. Seneviratne,
	R.~Cerci, J.~A. Lima, and R.~T. George, ``A stepwise approach to the visual
	interpretation of {CT}-based myocardial perfusion,'' \emph{Journal of
		Cardiovascular Computed Tomography}, vol.~5, no.~6, pp. 357--369, 2011.
	
	\bibitem{Ross14}
	A.~Rossi, D.~Merkus, E.~Klotz, N.~Mollet, P.~J. de~Feyter, and G.~P. Krestin,
	``Stress myocardial perfusion: imaging with multidetector {CT},''
	\emph{Radiology}, vol. 270, no.~1, pp. 25--46, 2014.
	
	\bibitem{Petr13b}
	R.~Petraco, S.~Sen, S.~Nijjer, M.~Echavarria-Pinto, J.~Escaned, D.~P. Francis,
	and J.~E. Davies, ``Fractional flow reserve--guided revascularization:
	practical implications of a diagnostic gray zone and measurement variability
	on clinical decisions,'' \emph{JACC: Cardiovascular Interventions}, vol.~6,
	no.~3, pp. 222--225, 2013.
	
	\bibitem{bahdanau2014neural}
	D.~Bahdanau, K.~Cho, and Y.~Bengio, ``Neural machine translation by jointly
	learning to align and translate,'' \emph{arXiv preprint arXiv:1409.0473},
	2014.
	
	\bibitem{he2015delving}
	K.~He, X.~Zhang, S.~Ren, and J.~Sun, ``Delving deep into rectifiers: Surpassing
	human-level performance on imagenet classification,'' in \emph{Proceedings of
		the IEEE international conference on computer vision}, 2015, pp. 1026--1034.
	
	\bibitem{coenen2018diagnostic}
	A.~Coenen, Y.-H. Kim, M.~Kruk, C.~Tesche, J.~De~Geer, A.~Kurata, M.~L. Lubbers,
	J.~Daemen, L.~Itu, S.~Rapaka \emph{et~al.}, ``Diagnostic accuracy of a
	machine-learning approach to coronary computed tomographic angiography--based
	fractional flow reserve: result from the {MACHINE} consortium,''
	\emph{Circulation: Cardiovascular Imaging}, vol.~11, no.~6, p. e007217, 2018.
	
	\bibitem{lu2017noninvasive}
	M.~T. Lu, M.~Ferencik, R.~S. Roberts, K.~L. Lee, A.~Ivanov, E.~Adami, D.~B.
	Mark, F.~A. Jaffer, J.~A. Leipsic, P.~S. Douglas \emph{et~al.}, ``Noninvasive
	{FFR} derived from coronary {CT} angiography: management and outcomes in the
	{PROMISE} trial,'' \emph{JACC: Cardiovascular Imaging}, vol.~10, no.~11, pp.
	1350--1358, 2017.
	
	\bibitem{Pontone2019}
	G.~Pontone, J.~R. Weir-McCall, A.~Baggiano, A.~D. Torto, L.~Fusini,
	M.~Guglielmo, G.~Muscogiuri, A.~I. Guaricci, D.~Andreini, M.~Patel,
	K.~Nieman, T.~Akasaka, C.~Rogers, B.~L. N{\o}rgaard, J.~Bax, G.~L. Raff,
	K.~Chinnaiyan, D.~Berman, T.~Fairbairn, L.~H. Koweek, and J.~Leipsic,
	``Determinants of rejection rate for coronary {CT} angiography fractional
	flow reserve analysis,'' \emph{Radiology}, p. 182673, jul 2019.
	
	\bibitem{Sakuma2019}
	H.~Sakuma, ``Coronary {CT} angiography fractional flow reserve analysis: A
	practical tool or a luxury for research centers?'' \emph{Radiology}, p.
	191044, jul 2019.
	
	\bibitem{jung2016patient}
	J.~Jung, Y.-H. Kim, N.~Kim, and D.~H. Yang, ``Patient-specific 17-segment
	myocardial modeling on a bull's-eye map,'' \emph{Journal of applied clinical
		medical physics}, vol.~17, no.~5, pp. 453--465, 2016.
	
\end{thebibliography}


\end{document}